\documentclass[pra,twocolumn,showpacs,preprintnumbers,amsmath,amssymb]{revtex4}

\usepackage{graphicx}
\usepackage{dcolumn}
\usepackage{bm}

\newcommand{\non}{\nonumber}
\newcommand{\mc}[1]{\mathcal{#1}}

\newcommand{\rr}{{\bf r}}

\newcommand{\ps}{\widehat{\psi}}
\newcommand{\ph}{\widehat{\phi}}
\newcommand{\aho}{a_{\rm HO}}

\begin{document}

\preprint{APS/123-QED}

\title{Vortex state in a superfluid Fermi gas near a Feshbach resonance}
\author{Yuki Kawaguchi}
\affiliation{Department of Physics, Tokyo Institute of Technology, Tokyo 152-8551, Japan}
\author{Tetsuo Ohmi}
\affiliation{Department of Physics, Kyoto University, Kyoto 606-8502, Japan}
\date{\today}

\begin{abstract}

We consider a single vortex in a superfluid Fermi gas in the BCS-BEC crossover regime near a Feshbach resonance.
The effect of the molecular Bose-Einstein condensate upon the vortex structure is discussed within the mean field approximation at zero temperature.
Using the self-consistent Bogoliubov-de Gennes equation of the fermion-boson coupled model,
we calculate density distributions of atoms and molecules.
As the number of the molecules increases, 
both atomic and molecular density changes from BCS-like distribution to BEC-like.
We also study the change of the vortex core size in the crossover regime.


\end{abstract}

\maketitle

\section{INTRODUCTION\label{sec:intro}}

The evidence of fermionic superfluidity in trapped atomic gases
has been clearly shown in the recent experiments~\cite{Regal2004,Zwierlein2004,Kinast2004,Bartenstein2004,Chin2004}.
The significant feature of these systems is that the strength of the inter-atomic interaction
is manipulated via a Feshbach resonance (FR) phenomena.
The crossover, which has been discussed several decades~\cite{Eagles1969,Legget1980,Nozieres1985},
between a weak coupling Bardeen-Cooper-Schrieffer (BCS) superfluid
and a Bose-Einstein condensate (BEC) of pre-formed pairs
can be realized in these systems.

The FR phenomena involve the scattering of atoms from open channel states
into a molecular bound state formed from neighboring closed channel states.
The so-called ``pre-formed pair'' in the conventional BCS-BEC crossover theory
is here equivalent to the substantial molecule in the closed channel,
which has been observed as a BEC on the one side of the FR.


In this paper we discuss the property of a single vortex in the BCS-BEC crossover regime at zero temperature.
Since the vortex states in BCS superfluids and in BECs are significantly differ in their distributions of particles,
observing a vortex state in a Fermi gas provides the information about superfluidity in the crossover regime.
In the case of a vortex in the BCS state, 
the energy gap is suppressed at the vortex core, and therefore
the particle-hole symmetrical modes, which are localized at the vortex core, exist in the vicinity of the Fermi surface~\cite{de-Gennes}.
Therefore, the atomic density is finite at the center of the vortex core.
On the other hand,
the particle density in a BEC vanishes at the vortex core, directly reflecting a singularity of the order parameter.

We are also interested in the vortex core size in the crossover regime.
The core size in a BCS superfluid is given by $\xi_{\rm BCS}\sim \hbar v_F/\Delta$,
where $v_F$ is the Fermi velocity and $\Delta$ is the energy gap.
As the attractive interaction becomes strong, the growth of the energy gap makes the core size smaller~\cite{Nygaard2003}.
As for a molecular BEC, the core size is written by $\xi_{\rm BEC}\sim \sqrt{1/8\pi n_M a_M}$,
where $n_M$ is the number density of molecules and $a_M$ is the {\it s}-wave scattering length of a molecule,
which diverges at resonance as well as that of an atom~\cite{Petrov2004, Randeria}.
So $\xi_{\rm BEC}$ also becomes small near the resonance.
Then, how does the core size change in the crossover regime?

There are several theoretical papers concerning vortex states in Fermi gases
~\cite{Rodriguez2001,Bruun-Viverit2001,Nygaard2003,Nygaard2004,Bulgac-Yu:condmat,Tempere:condmat}.
These papers, however, deal with only Fermi gases of which interaction strength is manipulated,
and the effect of a molecular BEC is not considered.
In this paper,
we start with an atom-molecule coupled model~\cite{Timmermans2001,Holland2001,Ohashi-Griffin2002},
and calculate density distributions and order parameters of both atoms and molecules.
We use the mean field approximation to simply deal with the effect of a molecular BEC.


\section{FORMALISM}
\label{sec:FORM}
We consider atomic gases of two atomic hyperfine states (labeled by $\sigma=\uparrow \downarrow$),
which are coupled to a molecular two-particle bound state.
The energy of a bare molecule relative to that of two bare atoms is denoted by $2\nu$.
The model Hamiltonian of the fermion-boson coupled system is given by
\begin{align}
\widehat{H}
=&\int d\rr \left[\sum_{\sigma} \ps_\sigma^\dagger(\rr)\mc{H}_A(\rr)\ps_\sigma(\rr)\right.\non\\
&+ \ph^\dagger(\rr)\left\{\mc{H}_M(\rr)+2\nu\right\}\ph(\rr) \non\\
&+\left. g\left\{\ph^\dagger(\rr)\ps_\downarrow(\rr)\ps_\uparrow(\rr) + {\rm h.c.}\right\}\right],
\label{eq:hamil}
\end{align}
where $\ps_\sigma$ and $\ph$ are the field operators of atoms and molecules, respectively.
The first and second terms in the integration represent the bare Hamiltonian of atoms and molecules, respectively, where
$\mc{H}_A(\rr)=-(\hbar^2/2m)\nabla^2+V_A(\rr)$, $\mc{H}_M(\rr)=-(\hbar^2/4m)\nabla^2+V_M(\rr)$,
$m$ is the atomic mass and $V_{A,M}(\rr)$ are the trapping potential for atoms and molecules, respectively.
The last term in Eq.~\eqref{eq:hamil} represents the atom-molecule coupling associated with FR.
Its contribution to the effective atom-atom interaction is $-g^2/2\nu$.
This means that the interaction is manipulated by changing $\nu$.
So we investigate the $\nu$ dependence of this system, especially at around $\nu=0$.
Although there exist non-resonant inter-atomic and intermolecular interactions,
we assume that the effect of the resonant process is so dominant that the other processes are negligible.

To conserve the total number of particles,
we use a single chemical potential $\mu$ and work with $\mc{H}=\widehat{H}-\mu \widehat{N}$,
where $\widehat{N}$ is the number operator given by
\begin{align}
\widehat{N}=\int d\rr \left[\sum_{\sigma}\ps_\sigma^\dagger(\rr)\ps_\sigma(\rr)+2\ph^\dagger(\rr)\ph(\rr) \right].
\end{align}

Here, we introduce the mean-field order parameters
$\mc{P}(\rr)\equiv\langle\ps_\downarrow(\rr)\ps_\uparrow(\rr)\rangle$ and
$\phi(\rr)\equiv\langle\ph(\rr)\rangle$,
which correspond to the Cooper pair amplitude and molecular order parameter, respectively.
The number density of molecules are given by $n_M(\rr)=|\phi(\rr)|^2$, as usual.
Though we neglect the bare atom-atom interaction,
the effective interaction via FR supports the pairing of atoms.
Since $\phi(\rr)$ is related to an energy gap as we will show below,
the pair amplitude remains finite as long as a molecular BEC exists.
The relation between these order parameters is given by
the equilibrium condition $i\hbar\langle \partial\ph(\rr)/\partial t\rangle=\langle[\ph(\rr),\mc{H}]\rangle=0$,
from which it follows that
\begin{align}
\left[\mc{H}_M(\rr)+2\nu-2\mu\right]\phi(\rr)+g\mc{P}(\rr)=0.
\label{eq:mol}
\end{align}

As in the case of the well-known mean field BCS theory,
the ``off-diagonal energy'' should be defined as the energy gap,
i.e., Eq.~\eqref{eq:hamil} implies that $\Delta_{\rm eff}(\rr)\equiv g\phi(\rr)$.
Then, the conventional Bogoliubov-de Gennes (BdG) approach is used in a similar manner:
the field operator of the atom is transformed with
\begin{eqnarray}
\left[ \begin{array}{c} \ps_\uparrow(\rr) \\ \ps_\downarrow^\dagger(\rr) \end{array} \right]
= \sum_j \left[
\begin{array}{cc} u_j^*(\rr) & -v_j(\rr) \\ v_j^*(\rr) & u_j(\rr) \end{array} \right]
\left[ \begin{array}{c} \alpha_{j\uparrow} \\ \alpha_{j\downarrow}^\dagger \end{array} \right],
\end{eqnarray}
and the BdG equation is obtained by
\begin{eqnarray}
\left[ \begin{array}{cc} \mc{H}_A(\rr)-\mu & g\phi^*(\rr) \\ g\phi(\rr) & -\mc{H}_A(\rr)+\mu \end{array} \right]
\left[ \begin{array}{c} u_j(\rr) \\ v_j(\rr) \end{array} \right]
= \epsilon_j \left[ \begin{array}{c} u_j(\rr) \\ v_j(\rr) \end{array} \right].
\label{eq:BdG}
\end{eqnarray}
In this notation, the pair amplitude and the density at zero temperature are written by
$\mc{P}(\rr)=-\sum_j u_j^*(\rr) v_j(\rr)$ and $n_A(\rr)=2\sum_j|v_j(\rr)|^2$, respectively.
To avoid the ultraviolet divergence in $\mc{P}(\rr)$,
we introduce a cutoff energy $\omega_c$ which is in the order of the Fermi energy,
and redefine $\mc{P}(\rr)=-\sum_j u_j^*(\rr) v_j(\rr)e^{-\epsilon_j^2/\omega_c^2}$.

Then, all we have to do is to solve Eqs.~\eqref{eq:mol} and \eqref{eq:BdG} self-consistently.
The chemical potential is determined so that the total number of atoms $N_{\rm tot}=N_A+2N_M$ is conserved,
where $N_{A,M}\equiv\int d\rr n_{A,M}(\rr)$. 


\section{Vortex state in the crossover regime}

We assume a cylindrical optical trap and approximate it with a two-dimensional harmonic potential:
$V_A(\rr)=(1/2)m\omega^2(x^2+y^2)$ and $V_M(\rr)=2V_A(\rr)$.
For simplicity, we neglect the $z$ dependence of the order parameters.
When considering the microscopic vortex structure, however,
the degree of freedom in $z$ direction is important.
So we assume that gases are confined in the length $L_z$,
and impose the periodic boundary condition.
We have numerically confirmed that the results does not depend on $L_z$ but on $N_{\rm tot}/L_z$
when $L_z\gtrsim\aho$, where $\aho=\sqrt{\hbar/2m\omega}$ is the harmonic oscillator length.

A single particle state of a bare atom in this potential is given by
$e^{ikz}e^{il\theta}(r/\aho)^l\mc{L}_n^{|l|}(r^2/2\aho^2)e^{-r^2/4\aho^2}$
having an energy eigenvalue $\epsilon_{kln}=\hbar\omega(2n+|l|+1+k^2\aho^2)$,
where $(r,\theta,z)$ is the cylindrical coordinate
and $\mc{L}_n^l(x)$ is the generalized Laguerre polynomial function.
The indices specifying the energy are given by
$l=0,\pm1,\pm2,\cdots$, $n=0,1,2,\cdots$, and
$k\aho=(2\pi\aho/L_z)n_k\equiv k_0 n_k$ where $n_k=0,\pm1,\pm2,\cdots$.
By counting the number of the eigenstates below $\epsilon_F$,
the Fermi energy as a function of the number of atoms is given by
$\epsilon_F(N_A)=\hbar\omega(15k_0 N_A/16)^{2/5}$.
The characteristic energy of the system is defined as $E_F\equiv \epsilon_F(N_{\rm tot})$,
which corresponds to the Fermi energy in the BCS limit.

For a vortex state in equilibrium, we assume the form
\begin{align}
\phi(\rr)&=e^{i\theta}\phi(r),\label{eq:phir}\\
u_{kln}(\rr)&=e^{ikz}e^{i(l-1)\theta}u_{kln}(r),\\
v_{kln}(\rr)&=e^{ikz}e^{il\theta}v_{kln}(r),
\end{align}
where $u_{kln}(r), v_{kln}(r)$ and $\phi(r)$ are real functions, 
and $n=0,1,2,\cdots$ is the radial quantum number.
In the BEC limit, by substituting $g\mc{P}=0$ in Eq.~\eqref{eq:mol}
the lowest energy state of a molecular BEC with a single vortex is given by
$\phi(\rr)=\sqrt{N_M/2\pi\aho^2L_z}e^{i\theta}f_0(r/\aho)$, where $f_0(r)\equiv \sqrt{2}re^{-r^2/2}$.
Here, the chemical potential is also determined by Eq.~\eqref{eq:mol} as $\mu=\nu+\hbar\omega$.

We consider a gas of $^6$Li atoms in a trap with $\omega=2\pi \times 300~{\rm Hz}$ and $\aho=1.7~\mu{\rm m}$.
The line density are set as $N_{\rm tot}/L_z=30~\mu$m$^{-1}$ and $200~\mu$m$^{-1}$,
leading to $E_F=9.7\hbar\omega$ and $21\hbar\omega$, respectively.
Though the particle density in our calculation is
about 10 times smaller than that in experiments,
$E_F/\hbar\omega$ are in the same order.
As for the coupling constant,
we consider a narrow FR and set $g\sqrt{n}=E_F$ and $3E_F$, where $n$ is the mean density of total atoms: $n=N_{\rm tot}/2\pi\aho^2L_z$.

\begin{figure}
\includegraphics[width=0.85\linewidth]{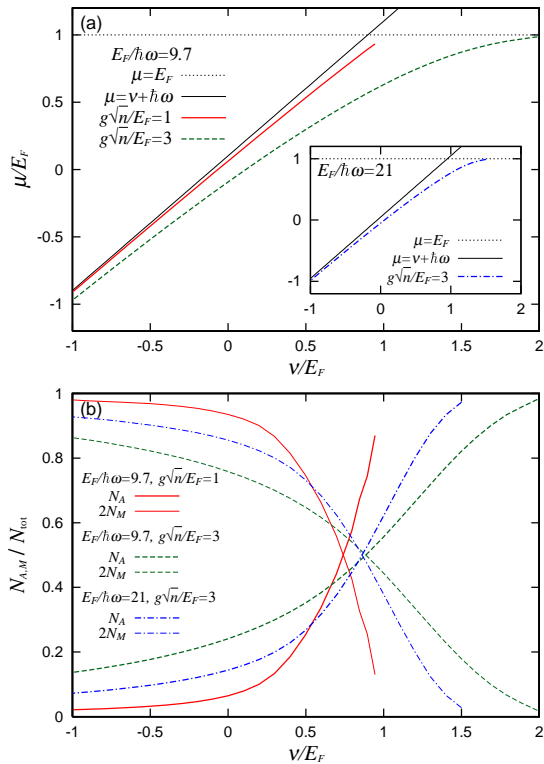}
\caption{(Color online) 
(a) The $\nu$ dependence of the chemical potential for $E_F=9.7\hbar\omega$.
The inset shows the results for $E_F=21\hbar\omega$.
The chemical potential far from the resonance goes to the asymptotic forms;
$\mu=E_F$ in the BCS limit ($\nu>0$) and $\mu=\nu+\hbar\omega$ in the BEC limit ($\nu<0$).
(b) The $\nu$ dependences of $N_A$ (thick lines) and $2N_M$ (thin lines).
The total density is always conserved: $N_A+2N_M=N_{\rm tot}$.
The intersections of $N_A$ and $2N_M$ depend upon $g\sqrt{n}/E_F$;
two intersections in the case of $g\sqrt{n}/E_F=3$ are coincide with each other.
}
\label{fig:1}
\end{figure}

Figure~\ref{fig:1}(a) shows the $\nu$ dependence of the chemical potential.
Since we neglect the bare atom-atom and molecule-molecule interactions,
the trapped gas behaves as an ideal gas far from the resonance in each side as shown in Fig.~\ref{fig:1}(a), i.e., 
in the BCS limit $\mu$ goes to $E_F$, while $\mu\to\nu+\hbar\omega$ in the BEC limit.
In Fig.~\ref{fig:1}(b), the numbers of atoms and molecules are plotted as a function of $\nu$.
It is clearly shown that
where the crossover occurs in the parameter space $\nu/E_F$ is determined by the interaction energy $g\sqrt{n}$ relative to $E_F$,
while the width of the crossover region depends on the coupling constant $g$.



\section{CORE STRUCTURE}
\begin{figure}
\includegraphics[width=0.85\linewidth]{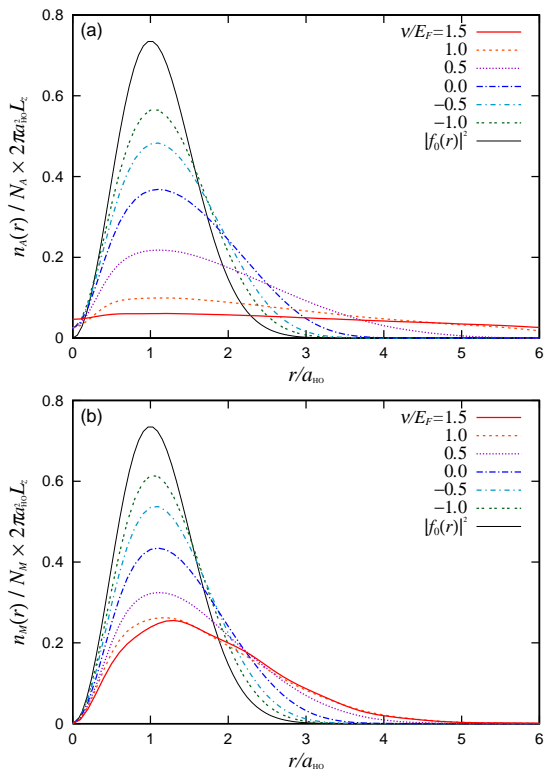}
\caption{(Color online) The density distributions of (a) atoms and (b) molecules in the case of $E_F=21\hbar\omega$ and $g\sqrt{n}=3E_F$.
The density profile of an ideal Bose gas is also plotted in both figures.
Both atoms and molecules change from broad distribution at $\nu/E_F=1.5$
to narrow one at $\nu/E_F=-1.0$.
The fraction of atoms localized at the core decreases as $\nu$ decreases,
while the molecular density at a vortex core is always zero.}
\label{fig:2}
\end{figure}

Figure~\ref{fig:2} shows the density distributions of (a) atoms and (b) molecules
in the case of $E_F=21\hbar\omega$ and $g\sqrt{n}=3E_F$.
Both profiles are normalized by the corresponding numbers of each particles.
We also plot the density distribution of a non-interacting BEC, $|f_0(r)|^2$, in both figures.
As we mentioned above, the atomic distribution far from resonance in BCS side ($\nu/E_F=1.5$)
has no explicit hole associated with a vortex.
The broad distribution in radial direction is also the feature of a fermionic profile.
As $\nu$ decreases, however, these fermionic features disappear and the atomic distribution becomes BEC-like, i.e.,
the cloud becomes narrower in radial direction and, moreover, the hole at the vortex core clearly appears.
This result is consistent with that of the one-channel model~\cite{Bulgac-Yu:condmat},
which argues that the density decreases at the vortex core when a BCS superfluid goes into the BEC regime.
Our calculation shows that this density depletion 
does not merely come from the increase of the molecular fraction.
The atomic profile itself also turns into a BEC-like distribution.

The existence of a molecular BEC leads to these changes in atomic distribution.
Since the molecular density always vanishes at vortex core as shown in Fig.~\ref{fig:2}(b),
the atomic density is also suppressed there when the number of molecules becomes large.
On the other hand, the molecular distributions in the BCS side are also strongly affected by atoms,
being broader than those in BEC side.
In the crossover regime, atoms and molecules are
strongly coupled to each other, and both gradually turn from BCS-like to BEC-like distribution.

\begin{figure}
\includegraphics[width=0.85\linewidth]{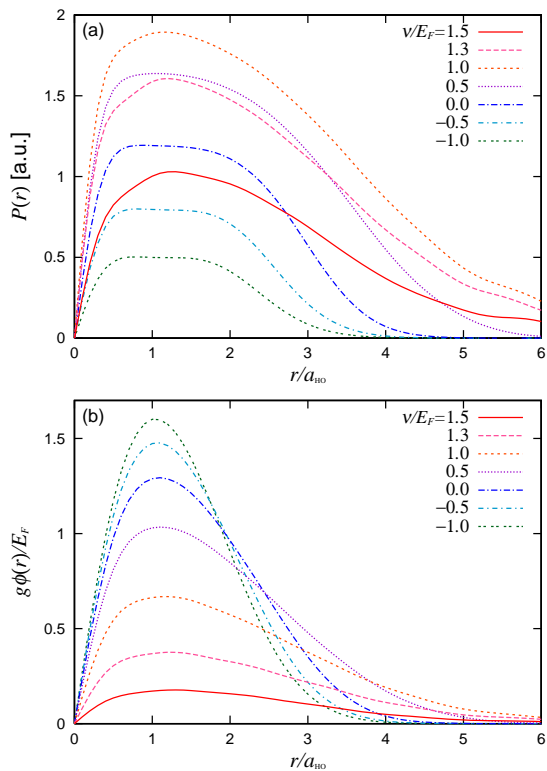}
\caption{(Color online) The spatial distribution of the mean field order parameters.
(a) The pair amplitude of atoms changes the $\nu$ dependence at $\nu/E_F\sim1.0$,
while (b) the molecular order parameter monotonically grows as $\nu$ increases.
}
\label{fig:3}
\end{figure}

Next, we discuss the change of order parameters.
Figure~\ref{fig:3} shows the $\nu$ dependence of
(a) the pair amplitude $\mc{P}(r)$ and (b) the molecular order parameter $\phi(r)$, or the effective energy gap.
These are also the results of the calculation with $E_F=21\hbar\omega$ and $g\sqrt{n}=3E_F$.
As $\nu$ decreases, the molecular order parameter monotonically grows and becomes narrower as well as the number density.
On the other hand, the fermionic pair amplitude becomes largest at $\nu/E_F=1.0$ and goes to zero in both BCS and BEC limit.
As $\nu$ decreases in the BCS side,
the pair amplitude increases since the effective energy gap, or $\phi(r)$, increases.
In the BEC side, however, an atomic pair turns into a molecule
and the pair amplitude becomes smaller as the number of molecules increases.
The $\nu$ dependence of the pair amplitude changes at $\nu/E_F=1.0$,
where the energy gap is equal to the chemical potential.

In Figs.~\ref{fig:3}(a) and (b), the order parameters have maxima at a nearly same radius.
This means that the core size does not strongly depend on $\nu$.
In the BCS side, the core region of $\mc{P}(r)$ becomes smaller as $\nu$ decreases (see, $\nu/E_F=1.5\sim1.0$ in Fig.~\ref{fig:3}(a)).
This fact is consistent with the BCS theory.
When the line density of total atoms is set larger as in the experiments,
leading to a broader distribution in radial direction,
this change must be clear.
In the region where $\nu/E_F<1.0$, however, the core size does not become smaller any more.
The core size in this region is determined by the scale of
spatial variations of the single-particle wavefunction with the energy $E_F$,
which is in the order of $\aho$.
In the case of free atoms, this size corresponds to the Fermi wave length $1/k_F\equiv\hbar/\sqrt{2mE_F}$.

As for the molecular order parameter,
the core region of $\phi(r)$ in BEC limit corresponds to the harmonic oscillator length
since we neglect the bare molecule-molecule interaction.
In the crossover regime, the atomic pair is expected to affect the core size of the molecular order parameter.
It seems, however, that the effect of the change in the $\mc{P}(r)$
on the core size in $\phi(r)$ is very little in our results.


\section{CONCLUSION}
We have studied a vortex state in a superfluid Fermi gas near a Feshbach resonance,
using the atom-molecule coupled model and the mean field approximation.
Coupled to each other, both atomic and molecular density distributions change from BCS-like to BEC-like in the crossover regime.
The depletion of the total particle density in BEC side
does not merely come from the increase of the number of molecules. 
Affected by a molecular BEC,
a hole at the core in atomic density clearly appears.
We have also studied the change of a vortex core size.
In the BCS side, the change of the core size is consistent with the BCS theory.
But becouse of the trapping potential, 
the core size dose not show drastic change in the crossover regime. 

\begin{acknowledgments}
The computation in this work has been done using
the facilities of the Supercomputer Center, 
Institute for Solid State Physics, University of Tokyo.
YK would like to acknowledge support from the Fellowship
Program of the Japan Society for Promotion of Science (Project No. 16-0648).
\end{acknowledgments}


\begin{thebibliography}{20}
\expandafter\ifx\csname natexlab\endcsname\relax\def\natexlab#1{#1}\fi
\expandafter\ifx\csname bibnamefont\endcsname\relax
  \def\bibnamefont#1{#1}\fi
\expandafter\ifx\csname bibfnamefont\endcsname\relax
  \def\bibfnamefont#1{#1}\fi
\expandafter\ifx\csname citenamefont\endcsname\relax
  \def\citenamefont#1{#1}\fi
\expandafter\ifx\csname url\endcsname\relax
  \def\url#1{\texttt{#1}}\fi
\expandafter\ifx\csname urlprefix\endcsname\relax\def\urlprefix{URL }\fi
\providecommand{\bibinfo}[2]{#2}
\providecommand{\eprint}[2][]{\url{#2}}

\bibitem[{\citenamefont{Regal et~al.}(2004)\citenamefont{Regal, Greiner, and
  Jin}}]{Regal2004}
\bibinfo{author}{\bibfnamefont{C.~A.} \bibnamefont{Regal}},
  \bibinfo{author}{\bibfnamefont{M.}~\bibnamefont{Greiner}}, \bibnamefont{and}
  \bibinfo{author}{\bibfnamefont{D.~S.} \bibnamefont{Jin}},
  \bibinfo{journal}{Phys. Rev. Lett.} \textbf{\bibinfo{volume}{92}},
  \bibinfo{pages}{040403} (\bibinfo{year}{2004}).

\bibitem[{\citenamefont{Zwierlein et~al.}(2004)\citenamefont{Zwierlein, Stan,
  Schunck, Raupach, Kerman, and Ketterle}}]{Zwierlein2004}
\bibinfo{author}{\bibfnamefont{M.~W.} \bibnamefont{Zwierlein}},
  \bibinfo{author}{\bibfnamefont{C.~A.} \bibnamefont{Stan}},
  \bibinfo{author}{\bibfnamefont{C.~H.} \bibnamefont{Schunck}},
  \bibinfo{author}{\bibfnamefont{S.~M.~F.} \bibnamefont{Raupach}},
  \bibinfo{author}{\bibfnamefont{A.~J.} \bibnamefont{Kerman}},
  \bibnamefont{and} \bibinfo{author}{\bibfnamefont{W.}~\bibnamefont{Ketterle}},
  \bibinfo{journal}{Phys. Rev. Lett.} \textbf{\bibinfo{volume}{92}},
  \bibinfo{pages}{120403} (\bibinfo{year}{2004}).

\bibitem[{\citenamefont{Kinast et~al.}(2004)\citenamefont{Kinast, Hemmer, Gehm,
  Turlapov, and Thomas}}]{Kinast2004}
\bibinfo{author}{\bibfnamefont{J.}~\bibnamefont{Kinast}},
  \bibinfo{author}{\bibfnamefont{S.~L.} \bibnamefont{Hemmer}},
  \bibinfo{author}{\bibfnamefont{M.~E.} \bibnamefont{Gehm}},
  \bibinfo{author}{\bibfnamefont{A.}~\bibnamefont{Turlapov}}, \bibnamefont{and}
  \bibinfo{author}{\bibfnamefont{J.~E.} \bibnamefont{Thomas}},
  \bibinfo{journal}{Phys. Rev. Lett.} \textbf{\bibinfo{volume}{92}},
  \bibinfo{pages}{150402} (\bibinfo{year}{2004}).

\bibitem[{\citenamefont{Bartenstein et~al.}(2004)\citenamefont{Bartenstein,
  Altmeyer, Riedl, Jochim, Chin, Denschlag, and Grimm}}]{Bartenstein2004}
\bibinfo{author}{\bibfnamefont{M.}~\bibnamefont{Bartenstein}},
  \bibinfo{author}{\bibfnamefont{A.}~\bibnamefont{Altmeyer}},
  \bibinfo{author}{\bibfnamefont{S.}~\bibnamefont{Riedl}},
  \bibinfo{author}{\bibfnamefont{S.}~\bibnamefont{Jochim}},
  \bibinfo{author}{\bibfnamefont{C.}~\bibnamefont{Chin}},
  \bibinfo{author}{\bibfnamefont{J.~H.} \bibnamefont{Denschlag}},
  \bibnamefont{and} \bibinfo{author}{\bibfnamefont{R.}~\bibnamefont{Grimm}},
  \bibinfo{journal}{Phys. Rev. Lett.} \textbf{\bibinfo{volume}{92}},
  \bibinfo{pages}{203201} (\bibinfo{year}{2004}).

\bibitem[{\citenamefont{Chin et~al.}(2004)\citenamefont{Chin, Bartenstein,
  Altmeyer, Riedl, Jochim, Denschlag, and Grimm}}]{Chin2004}
\bibinfo{author}{\bibfnamefont{C.}~\bibnamefont{Chin}},
  \bibinfo{author}{\bibfnamefont{M.}~\bibnamefont{Bartenstein}},
  \bibinfo{author}{\bibfnamefont{A.}~\bibnamefont{Altmeyer}},
  \bibinfo{author}{\bibfnamefont{S.}~\bibnamefont{Riedl}},
  \bibinfo{author}{\bibfnamefont{S.}~\bibnamefont{Jochim}},
  \bibinfo{author}{\bibfnamefont{J.~H.} \bibnamefont{Denschlag}},
  \bibnamefont{and} \bibinfo{author}{\bibfnamefont{R.}~\bibnamefont{Grimm}},
  \bibinfo{journal}{sience} \textbf{\bibinfo{volume}{305}},
  \bibinfo{pages}{1128} (\bibinfo{year}{2004}).

\bibitem[{\citenamefont{Eagles}(1969)}]{Eagles1969}
\bibinfo{author}{\bibfnamefont{D.~M.} \bibnamefont{Eagles}},
  \bibinfo{journal}{Phys. Rev.} \textbf{\bibinfo{volume}{186}},
  \bibinfo{pages}{456} (\bibinfo{year}{1969}).

\bibitem[{\citenamefont{Legget}(1980)}]{Legget1980}
\bibinfo{author}{\bibfnamefont{A.~J.} \bibnamefont{Legget}},
  \emph{\bibinfo{title}{{\rm in} Modern Trends in the Teory of Condensed
  Matter, {\rm ed. A. Pekalski} et. al.}}
  (\bibinfo{publisher}{Springer-Verlag}, \bibinfo{address}{Berlin},
  \bibinfo{year}{1980}), p.~\bibinfo{pages}{13}.

\bibitem[{\citenamefont{Nozi\`{e}res and Schmitt-Rink}(1985)}]{Nozieres1985}
\bibinfo{author}{\bibfnamefont{P.}~\bibnamefont{Nozi\`{e}res}}
  \bibnamefont{and}
  \bibinfo{author}{\bibfnamefont{S.}~\bibnamefont{Schmitt-Rink}},
  \bibinfo{journal}{J.\ Low\ Temp.\ Phys.} \textbf{\bibinfo{volume}{59}},
  \bibinfo{pages}{195} (\bibinfo{year}{1985}).

\bibitem[{\citenamefont{de~Gennes}(1989)}]{de-Gennes}
\bibinfo{author}{\bibfnamefont{P.~G.} \bibnamefont{de~Gennes}},
  \emph{\bibinfo{title}{Superconductivity of metals and alloys}}
  (\bibinfo{publisher}{Addison-Wesley}, \bibinfo{address}{New York},
  \bibinfo{year}{1989}), chap.~\bibinfo{chapter}{5}.

\bibitem[{\citenamefont{Nygaard et~al.}(2003)\citenamefont{Nygaard, Bruun,
  Clark, and Feder}}]{Nygaard2003}
\bibinfo{author}{\bibfnamefont{N.}~\bibnamefont{Nygaard}},
  \bibinfo{author}{\bibfnamefont{G.~M.} \bibnamefont{Bruun}},
  \bibinfo{author}{\bibfnamefont{C.~W.} \bibnamefont{Clark}}, \bibnamefont{and}
  \bibinfo{author}{\bibfnamefont{D.~L.} \bibnamefont{Feder}},
  \bibinfo{journal}{Phys. Rev. Lett.} \textbf{\bibinfo{volume}{90}},
  \bibinfo{pages}{210402} (\bibinfo{year}{2003}).

\bibitem[{\citenamefont{Petrov et~al.}(2004)\citenamefont{Petrov, Salomon, and
  Shlyapnikov}}]{Petrov2004}
\bibinfo{author}{\bibfnamefont{D.~S.} \bibnamefont{Petrov}},
  \bibinfo{author}{\bibfnamefont{C.}~\bibnamefont{Salomon}}, \bibnamefont{and}
  \bibinfo{author}{\bibfnamefont{G.~V.} \bibnamefont{Shlyapnikov}},
  \bibinfo{journal}{Phys. Rev. Lett.} \textbf{\bibinfo{volume}{93}},
  \bibinfo{pages}{090404} (\bibinfo{year}{2004}).

\bibitem[{\citenamefont{Randeria}(1995)}]{Randeria}
\bibinfo{author}{\bibfnamefont{M.}~\bibnamefont{Randeria}},
  \emph{\bibinfo{title}{{\rm in} Bose-Einstein Condensation, {\rm ed. A.
  Griffin} et. al.}} (\bibinfo{publisher}{Cambridge University Press},
  \bibinfo{address}{New York}, \bibinfo{year}{1995}), p. \bibinfo{pages}{355}.

\bibitem[{\citenamefont{Rodriguez et~al.}(2001)\citenamefont{Rodriguez,
  Paraoanu, and T\"{o}rm\"{a}}}]{Rodriguez2001}
\bibinfo{author}{\bibfnamefont{M.}~\bibnamefont{Rodriguez}},
  \bibinfo{author}{\bibfnamefont{G.-S.} \bibnamefont{Paraoanu}},
  \bibnamefont{and}
  \bibinfo{author}{\bibfnamefont{P.}~\bibnamefont{T\"{o}rm\"{a}}},
  \bibinfo{journal}{Phys. Rev. Lett.} \textbf{\bibinfo{volume}{87}},
  \bibinfo{pages}{100402} (\bibinfo{year}{2001}).

\bibitem[{\citenamefont{Bruun and Viverit}(2001)}]{Bruun-Viverit2001}
\bibinfo{author}{\bibfnamefont{G.~M.} \bibnamefont{Bruun}} \bibnamefont{and}
  \bibinfo{author}{\bibfnamefont{L.}~\bibnamefont{Viverit}},
  \bibinfo{journal}{Phys. Rev. A} \textbf{\bibinfo{volume}{64}},
  \bibinfo{pages}{063606} (\bibinfo{year}{2001}).

\bibitem[{\citenamefont{Nygaard et~al.}(2004)\citenamefont{Nygaard, Bruun,
  Schneider, Clark, and Feder}}]{Nygaard2004}
\bibinfo{author}{\bibfnamefont{N.}~\bibnamefont{Nygaard}},
  \bibinfo{author}{\bibfnamefont{G.~M.} \bibnamefont{Bruun}},
  \bibinfo{author}{\bibfnamefont{B.~I.} \bibnamefont{Schneider}},
  \bibinfo{author}{\bibfnamefont{C.~W.} \bibnamefont{Clark}}, \bibnamefont{and}
  \bibinfo{author}{\bibfnamefont{D.~L.} \bibnamefont{Feder}},
  \bibinfo{journal}{Phys. Rev. A} \textbf{\bibinfo{volume}{69}},
  \bibinfo{pages}{053622} (\bibinfo{year}{2004}).

\bibitem[{\citenamefont{Bulgac and Yu}(2004)}]{Bulgac-Yu:condmat}
\bibinfo{author}{\bibfnamefont{A.}~\bibnamefont{Bulgac}} \bibnamefont{and}
  \bibinfo{author}{\bibfnamefont{Y.}~\bibnamefont{Yu}},
  \bibinfo{journal}{cond-mat/0406256}  (\bibinfo{year}{2004}).

\bibitem[{\citenamefont{Tempere et~al.}(2004)\citenamefont{Tempere, Wouters,
  and Devreese}}]{Tempere:condmat}
\bibinfo{author}{\bibfnamefont{J.}~\bibnamefont{Tempere}},
  \bibinfo{author}{\bibfnamefont{M.}~\bibnamefont{Wouters}}, \bibnamefont{and}
  \bibinfo{author}{\bibfnamefont{J.~T.} \bibnamefont{Devreese}},
  \bibinfo{journal}{cond-mat/0410252}  (\bibinfo{year}{2004}).

\bibitem[{\citenamefont{Timmermans et~al.}(2001)\citenamefont{Timmermans,
  Furuya, Milonni, and Kerman}}]{Timmermans2001}
\bibinfo{author}{\bibfnamefont{E.}~\bibnamefont{Timmermans}},
  \bibinfo{author}{\bibfnamefont{K.}~\bibnamefont{Furuya}},
  \bibinfo{author}{\bibfnamefont{P.~W.} \bibnamefont{Milonni}},
  \bibnamefont{and} \bibinfo{author}{\bibfnamefont{A.~K.}
  \bibnamefont{Kerman}}, \bibinfo{journal}{Phys. Lett. A}
  \textbf{\bibinfo{volume}{285}}, \bibinfo{pages}{228} (\bibinfo{year}{2001}).

\bibitem[{\citenamefont{Holland et~al.}(2001)\citenamefont{Holland, Kokkelmans,
  Chiofalo, and Walser}}]{Holland2001}
\bibinfo{author}{\bibfnamefont{M.}~\bibnamefont{Holland}},
  \bibinfo{author}{\bibfnamefont{S.~J. J. M.~F.} \bibnamefont{Kokkelmans}},
  \bibinfo{author}{\bibfnamefont{M.~L.} \bibnamefont{Chiofalo}},
  \bibnamefont{and} \bibinfo{author}{\bibfnamefont{R.}~\bibnamefont{Walser}},
  \bibinfo{journal}{Phys. Rev. Lett.} \textbf{\bibinfo{volume}{87}},
  \bibinfo{pages}{120406} (\bibinfo{year}{2001}).

\bibitem[{\citenamefont{Ohashi and Griffin}(2002)}]{Ohashi-Griffin2002}
\bibinfo{author}{\bibfnamefont{Y.}~\bibnamefont{Ohashi}} \bibnamefont{and}
  \bibinfo{author}{\bibfnamefont{A.}~\bibnamefont{Griffin}},
  \bibinfo{journal}{Phys. Rev. Lett.} \textbf{\bibinfo{volume}{89}},
  \bibinfo{pages}{130402} (\bibinfo{year}{2002}).

\end{thebibliography}

\end{document}